%% file: FashionKE-demo.tex
\documentclass[sigconf]{acmart}

\AtBeginDocument{%
  \providecommand\BibTeX{{%
    \normalfont B\kern-0.5em{\scshape i\kern-0.25em b}\kern-0.8em\TeX}}}

\setcopyright{rightsretained} 

\begin{document}

\fancyhead{}

\copyrightyear{2019} 
\acmYear{2019} 
\acmConference[MM '19]{Proceedings of the 27th ACM International Conference on Multimedia}{October 21--25, 2019}{Nice, France}
\acmBooktitle{Proceedings of the 27th ACM International Conference on Multimedia (MM '19), October 21--25, 2019, Nice, France}\acmDOI{10.1145/3343031.3350607}
\acmISBN{978-1-4503-6889-6/19/10}

\title{Automatic Fashion Knowledge Extraction from Social Media}

\author{Yunshan Ma, Lizi Liao, Tat-Seng Chua}
\affiliation{
    \institution{National University of Singapore}
}
\email{yunshan.ma@u.nus.edu, liaolizi.llz@gmail.com, dcscts@nus.edu.sg}

\begin{abstract}
%Fashion knowledge plays a pivotal role in instructing people how to wear. 
Fashion knowledge plays a pivotal role in helping people in their dressing. In this paper, we present a novel system to automatically harvest fashion knowledge from social media. It unifies three tasks of occasion, person and clothing discovery from multiple modalities of images, texts and metadata.  A contextualized fashion concept learning model is applied to leverage the rich contextual information for improving the fashion concept learning performance. At the same time, to counter the label noise within training data, we employ a weak label modeling method to further boost the performance. We build a website to demonstrate the quality of fashion knowledge extracted by our system.
%We will demonstrate a website which can access and query the fashion knowledge extracted by our system.
\end{abstract}




\maketitle
\input{introduction}
\input{framework}
\input{implementation}
\input{conclusion}
\input{acknowledgement}

\bibliographystyle{ACM-Reference-Format}
\bibliography{mybib}

\end{document}

%% file: introduction.tex
\section{Introduction} \label{introduction}
Fashion is an integral part of human's daily life. Dressing well needs to consider not only the visual appearance of clothing, but also social conditions like the occasion and human identity. However, most existing works only focus on recognizing clothes, and few explore fashions at knowledge level. Fashion knowledge usually involves the interplay among three main aspects:  \textit{person}, \textit{clothing}, and \textit{occasion}. In general, social media sites such as Instagram provide a huge amount of user generated contents. A large portion of it contains people's up-to-date dressing codes involving the aforementioned interplay. It is indeed a natural and appropriate source to extract fashion knowledge.

%Clearly, there exists a large amount of patterns (\textit{e.g.}, dresscode or conventions) guiding people's daily fashion activities.

\begin{figure}[tp]
	\centering
	\includegraphics[scale=0.38]{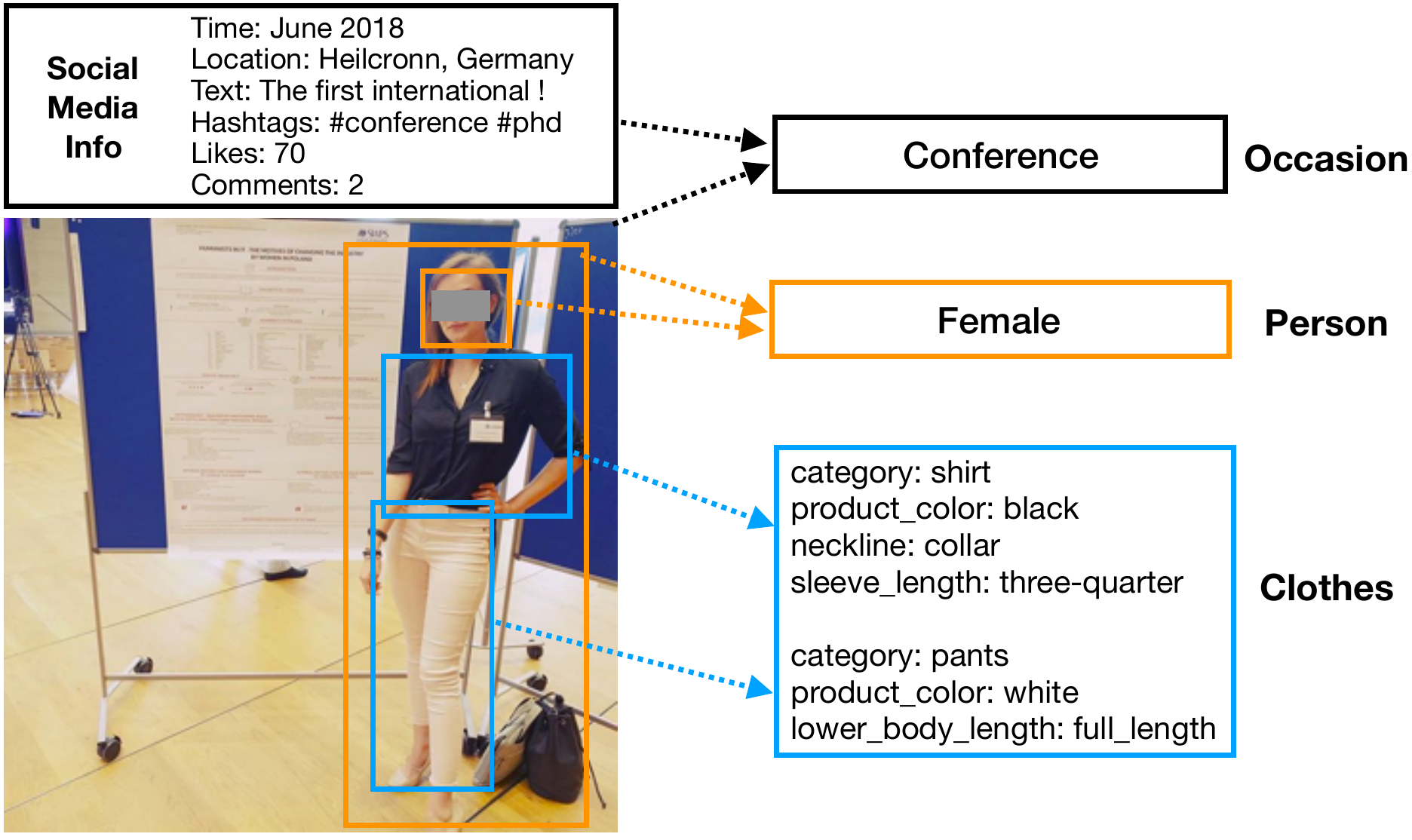}
	\vspace{-0.2cm}
	\caption{An illustration of fashion knowledge extraction from social media. It aims to extract triplets of <occasion, person, clothes> from multimodal inputs.}
	\label{Fig:illustration}
	\vspace{-0.4cm}
\end{figure}

In this paper, we devote to developing a system to automatically extract user-centric fashion knowledge from social media. An illustration of it %of fashion knowledge extraction from social media 
is shown in Figure \ref{Fig:illustration}. Here we aim to extract occasion, person, and clothing -- the three main aspects of fashion -- from social media posts which consist of multi-modal information like images, text, and metadata. However, fashion knowledge extraction from social media content is highly dependant on the performance of fashion concept prediction which has not been well addressed. Moreover, social media data lacks sufficient fashion concept labels which are crucial for fashion knowledge construction. To tackle these two problems, we first take advantage of the dependencies and correlations among different fashion concepts to improve the performance of fashion concept learning. Second, to alleviate the label insufficiency problem, we enrich the learning procedure with a weak label modeling module that utilizes both the machine-labeled and clean data. 

%% file: framework.tex
\section{Framework} \label{framework}

\begin{figure}[bp]
	\centering
	\vspace{-0.2cm}
	\includegraphics[scale=0.5]{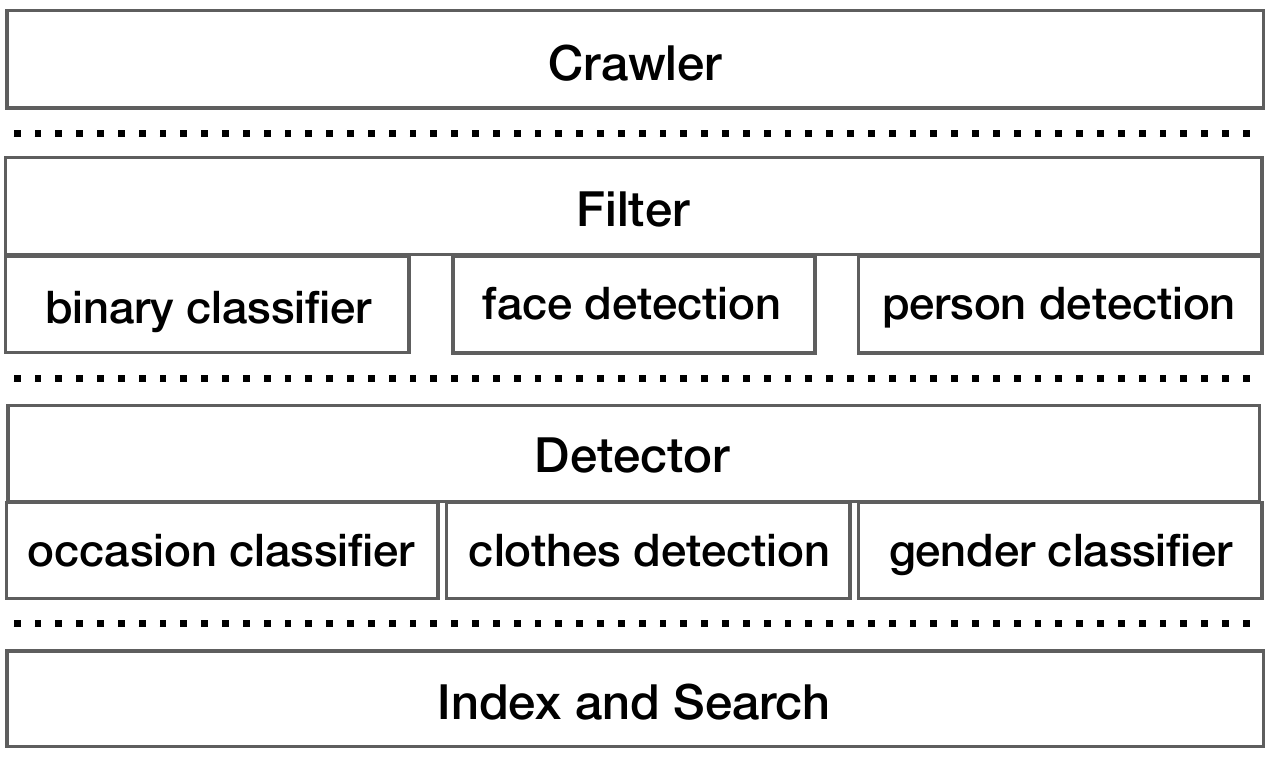}
	\vspace{-0.2cm}
	\caption{The overall architecture of our proposed system.}
	\label{Fig:pipeline}
	\vspace{-0.2cm}
\end{figure}

We present the architecture of our proposed system in Figure \ref{Fig:pipeline}. The first layer is the social media crawler. In the second layer, we employ multiple automatic filters to filter out posts that are not suitable to extract fashion concepts. Next, we employ fashion concept detection models to extract the fashion concepts. Lastly, we index all the extracted fashion knowledge and social media posts and build a website to access and query the fashion knowledge.

\begin{figure}[tp]
	\centering
	\vspace{-0.2cm}
	\includegraphics[scale=0.38]{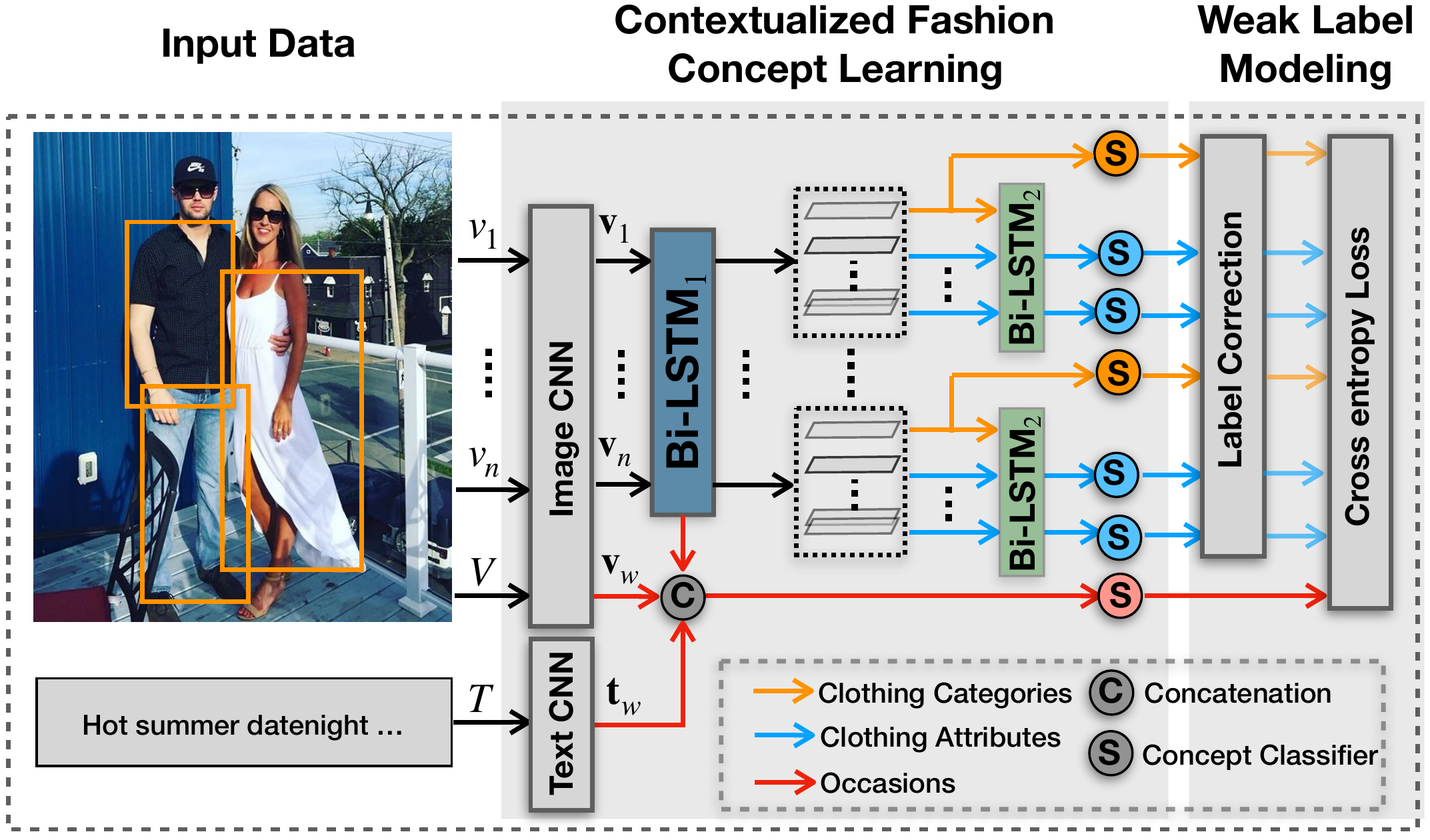}
	\vspace{-0.2cm}
	\caption{The structure of our proposed contextualized fashion concept learning model. Two bi-directional recurrent neural networks are used to model the correlation among different tasks, which improves the performance of fashion concept learning. A learned label transition matrix is utilized to counter the noise in machine-labeled data.}
	\label{Fig:framework}
	\vspace{-0.5cm}
\end{figure}

Specifically, the crawler gets the images and side information, and inputs them into the filters. After filtering, the remaining high-quality posts are fed into detectors to extract fashion concepts related to occasion, person, and clothes. Finally, the fashion concept triplets and the associated social media information are organized into a knowledge base for further fashion knowledge search.  

Among the four layers of this architecture, the most crucial part is to detect the fashion concepts from the posts. Figure \ref{framework} shows our detection model. (1) We design a contextualized fashion concept learning model, in which two bi-directional recurrent neural networks are utilized to model co-occurrence among occasion, clothing attributes and categories. %Specifically, we use $BiLSTM_1$ to capture the dependency among different clothes areas, contributing better representations for each clothes region. The $BiLSTM_2$ is designed to model the correlation among different attributes of each clothes, which may affect each other. 
(2) We introduce a weak label modeling module 
%which estimates a label transition matrix for bridging the gap between weak labels and clean labels. It enables 
to leverage the large amount of cheap machine-labeled training data. For more details, please refer to \cite{ma2019fashionKE}.

After extracting the fashion concepts, we construct a triplet (<occasion, person, clothes>) for each cloth, which is the candidate of fashion knowledge we aim to harvest. We aggregate all the triplets and instances to construct a fashion knowledge base. We design a website as an interface to access and query the extracted fashion knowledge extracted by our system. As shown in Figure \ref{Fig:frontend}, the well-organized repository provides a useful platform for us to query and discover fashion knowledge and trends.

%% file: implementation.tex
\section{Implementation} \label{implementation}

\begin{figure}[!htp]
	\centering
	\includegraphics[scale=0.35]{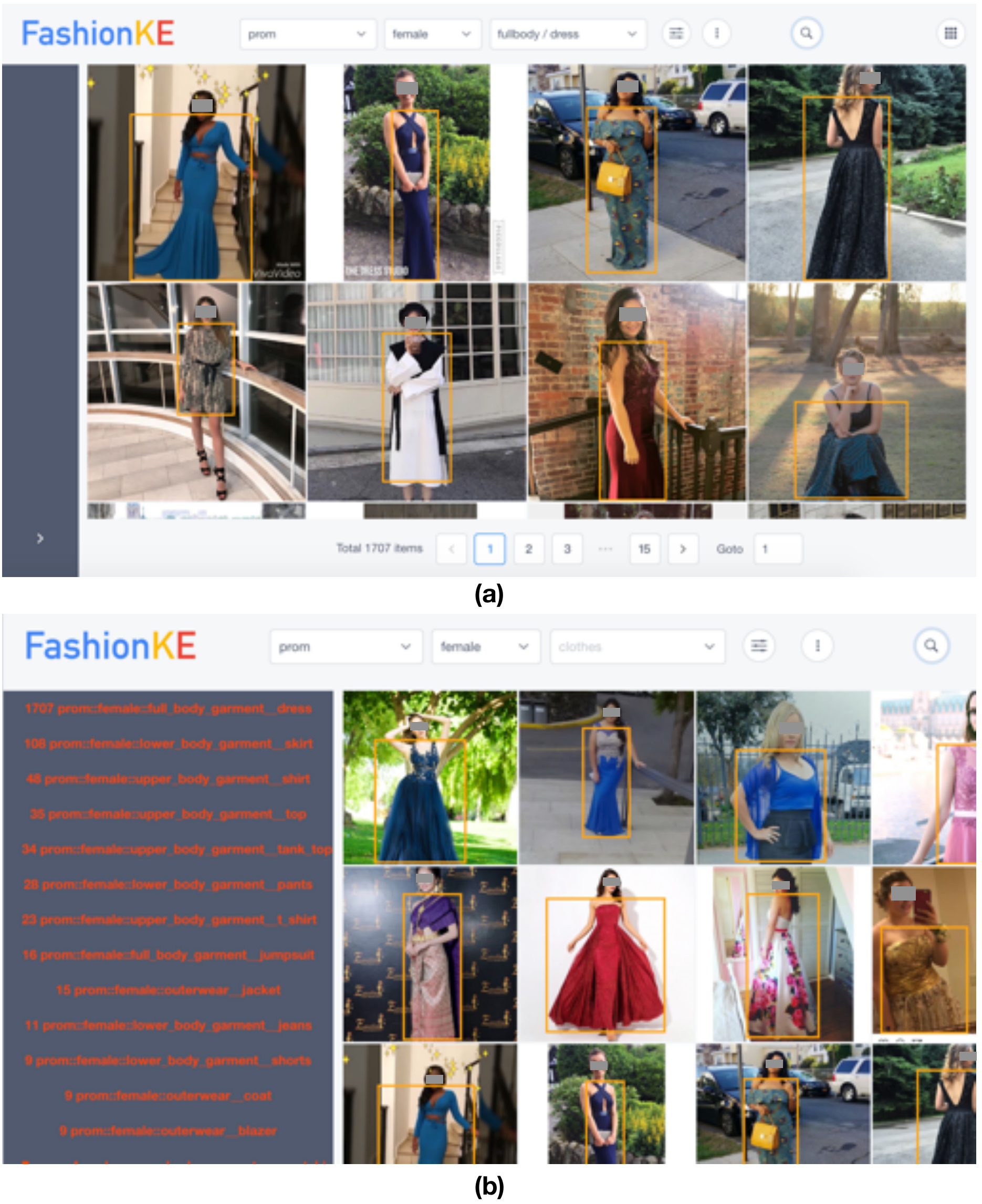}
	\vspace{-0.2cm}
	\caption{The interface of the website to explore the extracted fashion knowledge. (a) illustrates the images which contain the triplet: <prom, female, dress>; and (b) the left side shows all the triplets (ordered by the number of instances of each triplet) that have occasion prom and gender female.}
	\label{Fig:frontend}
	\vspace{-0.3cm}
\end{figure}

\textbf{Crawler and Filter}. We crawl Instagram posts by hashtags, which are manually chosen and cover 10 occasions. However, a large portion of the posts are unrelated to fashion concepts. Thus, we first detect the faces and bodies in the images and remove those images which do not have any face-body pair. To ensure the visibility of clothes, we only keep those images with proper face-body-image ratios ($\frac{height\_of\_face}{height\_of\_body} < 0.2$ and $\frac{height\_of\_body}{height\_of\_image} > 0.5$). Moreover, we train a binary classifier on images to help us remove those images of posters, news or advertisements. % Since everyday there are large amount of fresh posts, our crawler works $24\times7$ continuously and follows the latest fashion trends.

\textbf{Detector}. We first use a pre-trained object detection model to detect the clothes (bounding boxes and rough categories) in the images. We then implement the contextualized fashion concept learning model to predict the occasions, categories and attributes of clothes. In this paper, we use 10 occasions, 21 clothes categories, and 8 clothes attributes with 50 attribute values. A pre-trained gender prediction model is used to predict the gender.

\textbf{Index and Search}. We first construct our fashion knowledge base based on the fashion triplets and the associated posts. We index not only the occasion, gender, and clothes, but also texts and metadata, such as time, location, likes, and comments. Based on such a data backbone, we design a website to serve as an interface between users and backend data. Figure \ref{Fig:frontend} shows the frontend of our website, which supports two types of the search: triplets and posts. In terms of search options, we support both fashion concept level options (occasion, gender, clothes category and attributes) and metadata level options (time, location, hashtag, likes, comments).   

%% file: conclusion.tex
\section{Conclusion} \label{conclusion}
In this paper, we constructed a system to automatically extract fashion knowledge from social media. We demonstrated the system, presented the techniques employed, and shared our experience in dealing with the challenges in this field. Several points could be further explored in future to improve the fashion knowledge extraction system including trending analysis of different fashion concepts and influence diffusion analysis among social networks.

%% file: acknowledgement.tex
\section*{acknowledgement}
This research is part of NExT++ project, which is supported by the National Research Foundation, Prime Minister's Office, Singapore under its IRC@SG Funding Initiative.